\documentclass[reprint,superscriptaddress,amsmath,amssymb,aps,prd]{revtex4-2}

\usepackage[colorlinks=true, allcolors=blue]{hyperref}
\usepackage{xcolor}
\usepackage{romannum}
\usepackage{graphicx}
\usepackage{dcolumn}
\usepackage{orcidlink}
\usepackage{tabularx}
\usepackage{longtable}
\usepackage{booktabs}

\newcommand{\ba}{\begin{align}}
\newcommand{\ea}{\end{align}}

\begin{document}

\preprint{\textbf{???}}

\title{Searching for new physics with contact-free transitions in muonic atoms}

\author{Noam~Burger\orcidlink{0009-0009-6347-5672}}
\affiliation{Physics Department, Technion—Israel Institute of Technology, Haifa 3200003, Israel}
\affiliation{The Helen Diller Quantum Center, Technion—Israel Institute of Technology, Haifa 3200003, Israel}

\author{Ben~Ohayon\orcidlink{0000-0003-0045-5534}}
\email[Corresponding author: ]{bohayon@technion.ac.il}
\affiliation{Physics Department, Technion—Israel Institute of Technology, Haifa 3200003, Israel}
\affiliation{The Helen Diller Quantum Center, Technion—Israel Institute of Technology, Haifa 3200003, Israel}

\date{\today}

\begin{abstract}

We present a feasibility study demonstrating how recent advances in quantum-sensing technologies can facilitate stringent comparisons between theoretical predictions and experimental measurements of \emph{contact-free} transitions in muonic atoms.
Such comparisons can probe previously unexplored regions of the parameter space governing spin-independent muon-proton interactions and disentangle the extraction of fundamental constants from searches for physics beyond the Standard Model.
\end{abstract}

\maketitle

\section{\label{Introduction}Introduction}

The physics community is engaged in a global effort to find phenomena beyond the Standard Model. Such effects may appear as significant differences between measured observables (e.g., the spacings of atomic energy levels) and theoretical predictions.
%
In the nonrelativistic limit, a new spin-independent interaction manifests as an effective Yukawa potential~\cite{yukawa1935interaction}
\begin{align}
V_{X}(r) = (-1)^{s}\frac{g_1^X g_2^X}{4\pi}\frac{e^{-m_X r}}{r},
\label{eq:Yukawa potential}
\end{align}
where $X$ denotes a new boson of spin $s$, mass $m_X$, and coupling constants $g_1^X$ and $g_2^X$ to the bound particles. 
At leading order, and for circular states ($l=n-1$), the resulting shift of the $n$th energy level can be expressed as \cite{liu2025probing}
\begin{align}
\Delta E_n^X = (-1)^{s}\frac{g_1^X g_2^X}{4\pi} \frac{1}{r_n} \frac{1}{\left(1+\frac{m_X r_n}{2 n}\right)^{2n}}\: ,
\label{eq:NP energy change}
\end{align}
where 
\begin{align}\label{eq:dec}
r_n=n^2/(Z\alpha\,m_r)
\end{align}
is a decoupling parameter inversely proportional to the reduced mass $m_r$ and the binding parameter $Z\alpha$; consequently, bound systems where both constituents are heavier than the electron, such as muonic atoms, probe new physics associated with heavier mediators.

Different experimental systems probe various combinations of couplings and mass ranges (see Ref.~\cite{delaunay2026atomic} and references therein). In previous work, we explored hadronic atoms, which probe couplings to protons and neutrons~\cite{liu2025probing}.
Here, we focus on muonic atoms, which, at tree level, are sensitive to interactions involving muons and nucleons: $g_\mu^X g_p^X$ and $g_\mu^X g_n^X$.
\begin{figure}[tb]
    \centering
    \includegraphics[width=0.99\linewidth, trim=0 3mm 0 2mm, clip]{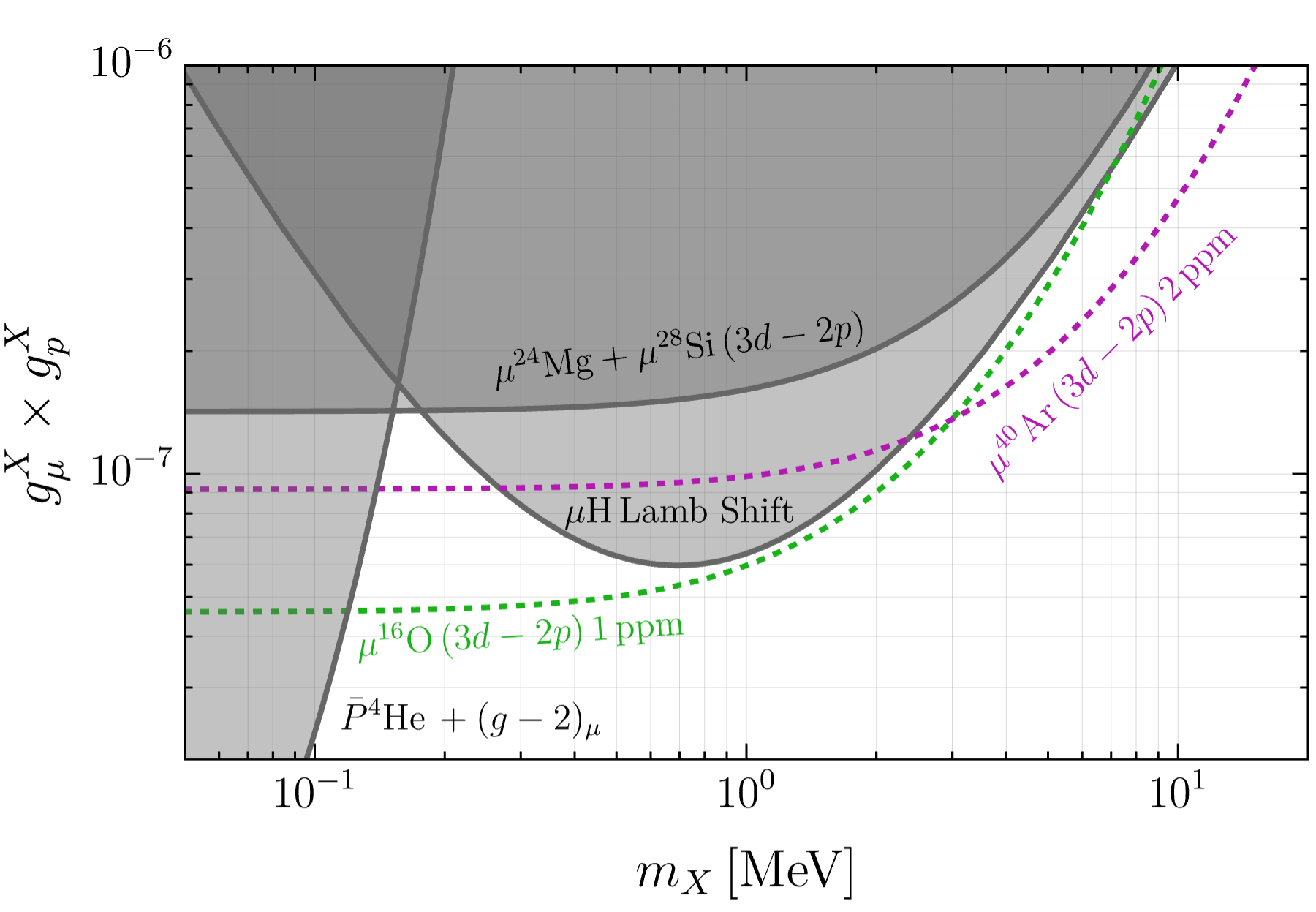}
    \caption{
    Current bounds on $g_\mu^X g_p^X$, and the sensitivity of the measurements whose feasibility is explored in this work.
    See main text for details.
    }
    \label{fig:bounds on new physics}
\end{figure}
%
Existing constraints on $g_\mu^X g_n^X$ are stringent.
For a new boson with $m_X \lesssim 20 \, \text{MeV}$, they arise from a combination of the muon anomalous magnetic moment $(g-2)_\mu$\,\cite{2025-g-2} and neutron scattering experiments~\cite{leeb1992, 2017-Scattering}. See however Ref.~\cite{2021-custodial}, which shows relevant new physics scenarios that do not affect the muon $(g-2)_\mu$.
For heavier mediators, the leading constraints come from the combination of muonic and electronic isotope shifts between hydrogen and deuterium~\cite{2011-HDIS, 2017-Scattering, pohl2016laser}.
The stringency of existing bounds on $g_\mu^X g_n^X$ makes absolute spectroscopy measurements in muonic atoms primarily sensitive to $g_\mu^X g_p^X$.
Accordingly, this work focuses on the prospects for probing unexplored regions of the $g_\mu^X g_p^X$ vs. $m_X$ parameter space. To do so, one must compare experimentally measured muonic atom transition energies with their theoretical predictions at an unprecedented level of accuracy.

Existing bounds on $g_\mu^X g_p^X$ for $50\,\text{keV}<m_X$ are shown in Fig.~\ref{fig:bounds on new physics}.
Up to a mass of $0.15 \, \text{MeV}$, they arise from the multiplication of $g^X_\mu<10^{-4}$ (combining Eq.~(2.3) of Ref.~\cite{2018-au} with Eq.~(1.5) of Ref.~\cite{2026-au}) and $g^X_p$ from antiprotonic helium spectroscopy~\cite{doi:10.1126/science.aaf6702}.
For heavier mediators, the values arise from Lamb Shift measurements in muonic atoms using charge radii obtained from a fit that determines fundamental constants with non muonic atom data (see table \Romannum{16} of Ref. \cite{mohr2025codata}). The most stringent bound is from $\mu $H~\cite{pohl2010size,antognini2013proton}. It is limited by the extraction of the proton radius from electronic measurements and by the tension between experiment and theory, referred to as the \textit{proton radius puzzle}~\cite{2013-Puzzle}.

The requirement for precisely determined radii from electronic systems can be alleviated by measuring transitions that do not involve $s$- or $p_{1/2}$-states, which we refer to as \emph{contact-free}.
As shown in Eq.~(\ref{eq:dec}), shifting to levels with high principal quantum numbers decreases sensitivity to heavy new bosons.
Moreover, transitions between these levels involve a large number of closely spaced fine-structure components, complicating the extraction of individual line positions and increasing the associated uncertainties.
As we will show below, transitions in the $3d-2p_{3/2}$ manifold best balance experimental accessibility and theoretical simplicity. Notice that the $3d$ fine structure is too small to be resolved. 
In the past, measurements of these transitions in $\mu^{24}\text{Mg}$ and $\mu^{28}\text{Si}$ achieved a precision of $4$~ppm~\cite{beltrami1986new}, which, as shown in Fig.~\ref{fig:bounds on new physics}, represents the upper limit for further exploration of the new physics parameter space.

In this work, we present a strategy to exploit advancements in both experiment and theory to confront experimental results with theoretical predictions in contact free transitions of muonic atoms with nuclear charges up to $Z=20$.
The target accuracy is up to $1~$ppm, which explores new regions of the physics parameter space and enables a complete decoupling of the proton (and deuteron) radius puzzle from interpretations of new physics.

\section{\label{Feasability}Feasibility study}

\subsection{\label{Sec: QED} QED calculations}

\begin{figure}[tb]
    \centering
    \includegraphics[width=0.99\linewidth, trim=0 4mm 0 2mm, clip]{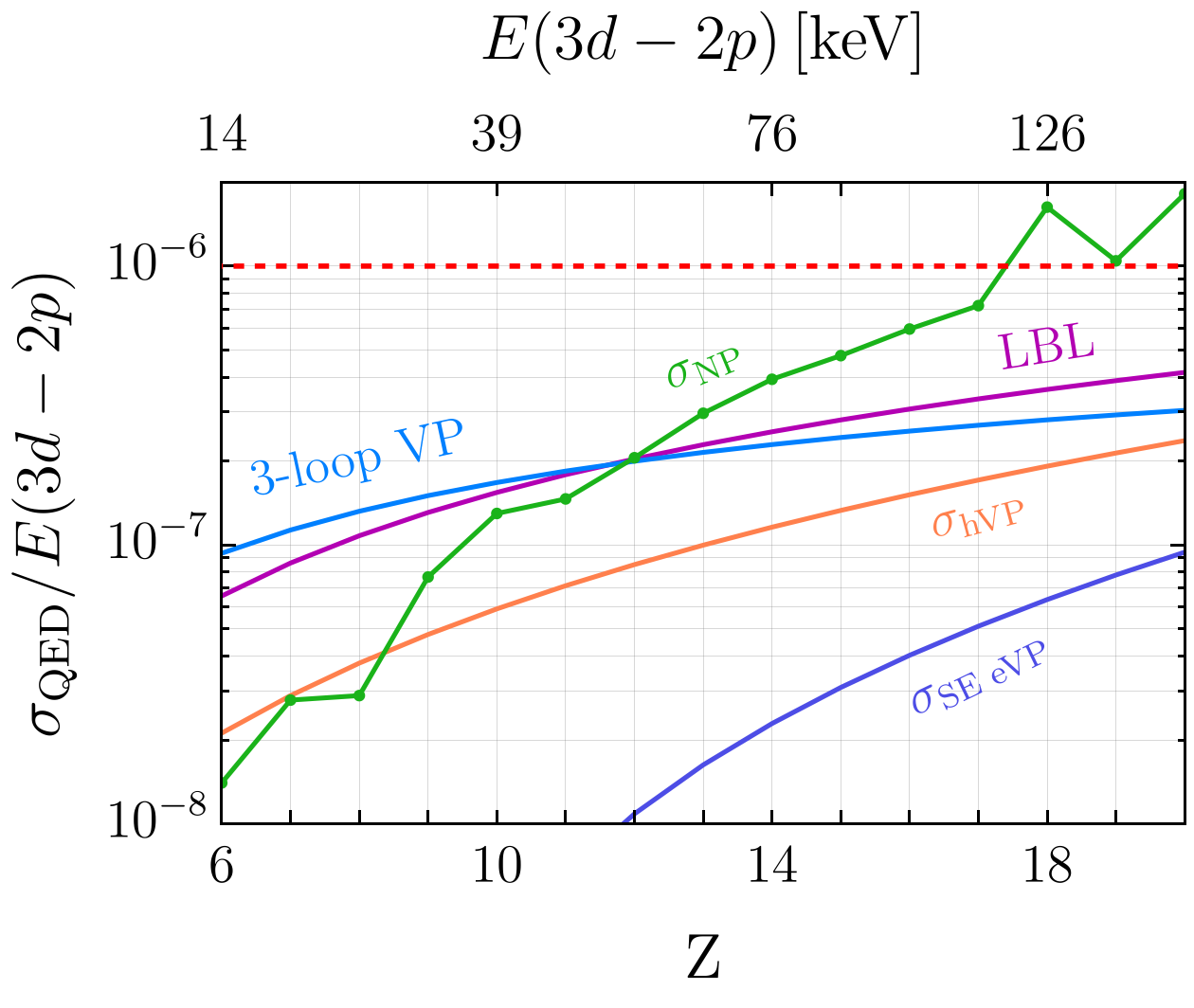}
    \caption{Estimated relative uncertainty contributions to the theoretical calculation of the centroid energy of the $3d-2p$ manifold from not-yet-calculated QED corrections: three loop vacuum polarization - VP and light-by-light scattering - LBL, and uncertainties in existing values: nuclear polarization - NP, hadronic vacuum polarization - hVP, and the self-energy combined with electronic VP - SE-eVP.
    The top axis indicates the centroid energy of the $3d-2p$ manifold for a certain $Z$.}
    \label{fig:QED and NP uncertainty}
\end{figure}

The bound-state QED theory of contact-free transitions in light muonic systems is considered accurate to better than $1$\,ppm~\cite{2016-PiMass, 2023-NePRL, 2026-ubarspec, 2026-BeTh}, excluding the nuclear structure effects discussed next. However, several higher-order QED terms for the systems under consideration remain uncalculated and contribute to the theoretical uncertainty. Their order of magnitude is estimated in this section and plotted as a function of $Z$ in Fig.~\ref{fig:QED and NP uncertainty}.

The first missing correction we consider is the three-loop electronic vacuum polarization (eVP), which has only been evaluated for the lightest systems~\cite{RevModPhys.96.015001, 2017-MArty}. We follow Ref.~\cite{2026-ubarspec} and estimate the uncertainty from the  exclusion of three-loop eVP as $\alpha^2$ times the leading eVP contribution.
Another missing contribution is the light-by-light (LBL) scattering correction~\cite{Karshenboim:2010cq}. To estimate its magnitude, we first calculate the Wichmann--Kroll contribution (using the public code of Ref.~\cite{2026-ubarspec}) and then divide the result by $Z$. 
The result grows rapidly with nuclear charge, making it the leading contribution above $Z=12$.
An additional uncertainty arises from the hadronic vacuum polarization (hVP) correction.
This contribution can be obtained at leading order by scaling the muonic VP correction by a constant factor equal to $\gamma_{had}=0.6746(160)$~\cite{RevModPhys.96.015001}.
The uncertainty on this factor is propagated to obtain the uncertainty on the hVP correction.
An additional uncertainty arises from the electronic vacuum polarization correction to the self-energy (SE-eVP). It was calculated in Ref.~\cite{Ohayon:2024akb} up to $Z=6$. Extrapolating to higher Z with an uncertainty of $100\%$ results in a small uncertainty contribution, even for $Z=20$.
Finally, we note that the dependence of the relativistic-recoil correction in muonic atoms on nuclear shape requires further theoretical efforts~\cite{2025-RecoilFNS}.

The quadratic sum of all the considered uncertainties reaches $0.5\,$ppm for $Z=20$. This value is small enough that pure QED effects are not expected to hinder new-physics searches within our targeted accuracy; however, any enhancement beyond our estimates could make it significant, justifying additional investigation by the community.

\subsection{\label{Sec: Screen} Nuclear Polarization}

A major advantage of contact-free transitions is the simplification and reduction of nuclear polarization (NP) effects.
For circular states with $Z\alpha/n \ll 1$, the leading order energy shift can be expressed as 
\begin{align}
E_n^{\mathrm{NP}} = \alpha_d h_n^{\mathrm{NP}},
\label{eq:NP}
\end{align}
where $h_n^\mathrm{NP}\equiv-8m_r^4Z^4\alpha^5(2n-4)!/\left(n^4(2n)!\right)$ and $\alpha_d$ is the static electric dipole polarizability~\cite{2024-P, liu2025probing, 2025-EFT}. 
Recent work obtained $\alpha_d$ from an artificial neural network optimized for analyzing photoabsorption cross section data in light nuclei~\cite{2026-alphad}. The reported uncertainties arise from variations among accepted networks under fixed methodological choices and should be considered lower limits. For other nuclei of interest, $\alpha_d$ is taken with $10\%$ uncertainty from Ref.~\cite{2026-NPMIsha}.
The uncertainty associated with this contribution is shown in Fig.~\ref{fig:QED and NP uncertainty}. 
Its increasing magnitude with $Z$ supports the selection of light elements ($Z<17$) for the most stringent accuracy goals. It also motivates the community to complement the data-driven values of $\alpha_d$ with those calculated \textit{ab initio} and to develop NP theory for contact-free states beyond leading order.

\subsection{\label{Sec: Screen} Nuclear size and shape}
At leading order in $\alpha Z$, the finite nuclear size affects only the $2p_{3/2}$ state and is proportional to the fourth power of the fourth moment of the charge distribution, $r_{CC}^4$~\cite{pachucki2018three}.
It is beneficial to consider $r_{CC}$ as a product of the RMS charge radius $r_C$ and a dimensionless nuclear shape parameter $r_{CC}/r_C$~\cite{2026-LiLike, 2026-FS}. Unlike $s$ states, $p_{3/2}$ states probe both size and shape with comparable sensitivity.

For nuclei with $Z>2$, values of $r_c$ are primarily determined from $2p-1s$ transitions in muonic atoms~\cite{ohayon2025critical}. These may themselves be affected by new physics. However, the experimental precision in the energies is worse than $5~\,$ppm~\cite{1980-BeBN, 1981-Wohlfahrt, 1985-12C, 1985-SP, 1992-Fricke, 2026-Cl, Eizenberg2026Be, deseyn2026}, so any possible beyond-Standard-Model contributions are negligible given the existing bounds.
The recently updated charge radii~\cite{ohayon2025critical, Eizenberg2026Be, 2026-Cl, 2025-Testing, 2026-Ar} result in relative uncertainties for $3d-2p$ transitions below $0.4$\,ppm for all considered nuclei, except for $^{40}$Ca, where it is $0.6$\,ppm. Therefore, the uncertainty from nuclear size is mostly negligible. 

\begin{figure}[tbp]
    \centering
    \includegraphics[width=0.99\linewidth, trim=0 3mm 0 6mm, clip]{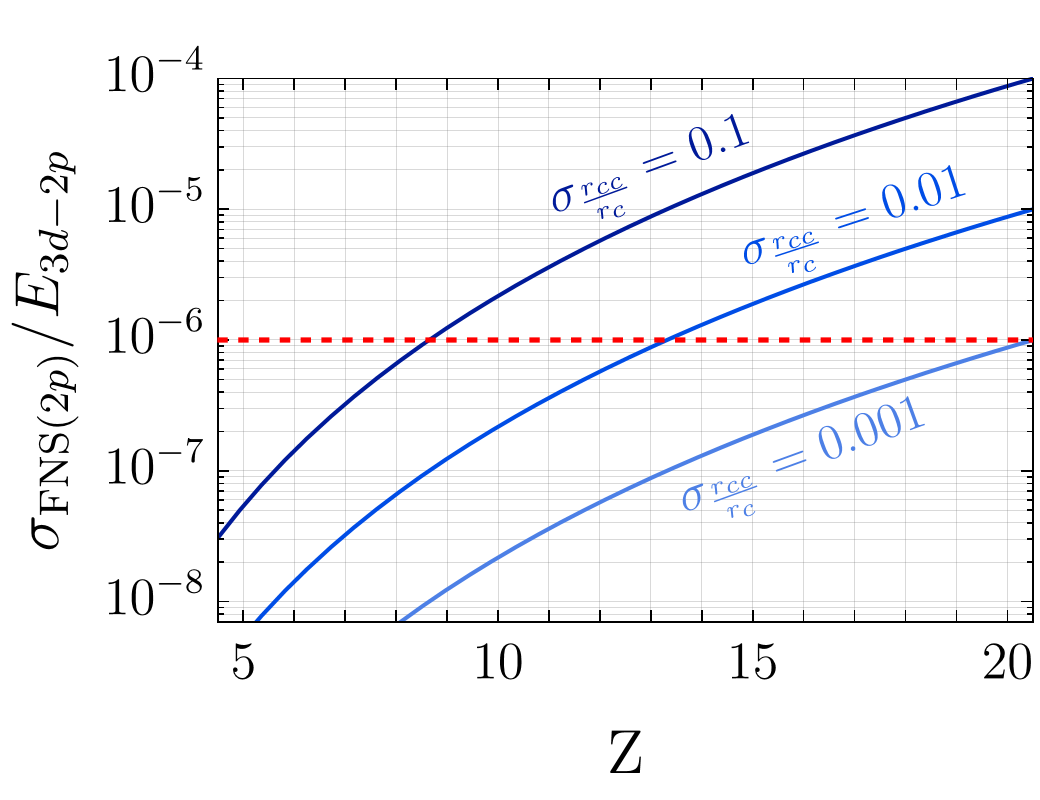}
    \caption{Relative uncertainty of the finite-nuclear-size (FNS) correction to the $2p_{3/2}$ state as a function of the atomic number $Z$. The three blue curves correspond to different assumptions for the uncertainty in the nuclear-shape $r_{CC}/r_C$ as detailed in the main text. The dashed red line indicates a relative uncertainty of $1$ ppm.}
    \label{fig:FNS uncertainty}
\end{figure}

Because $r_C$ is much better understood than $(r_{CC}/r_C)$, the uncertainty in the nuclear shape correction dominates.
In Fig.~\ref{fig:FNS uncertainty}, we plot it as a function of $Z$ under varying assumptions about the uncertainty in $(r_{CC}/r_C)$.
A relative shape parameter uncertainty of $10\%$ in light nuclei was estimated from the difference in $(r_{CC}/r_C)$ between Gaussian and exponential charge distributions~\cite{2017-MArty}. This outcome would be unfavorable for the current proposal, as shown in Fig.~\ref{fig:FNS uncertainty}; however, it is likely too conservative.
Ongoing work by the authors suggests that in these light systems, $(r_{CC}/r_C)$ can be reliably extracted from charge distributions derived from elastic electron scattering experiments with an accuracy on the order of $1\%$. The corresponding uncertainty contribution, shown in Fig.~\ref{fig:FNS uncertainty}, suggests focusing on $Z<14$, where it remains smaller than $1\,$ppm. 
To better determine $r_{CC}/r_C$ and enable competitive new physics searches in heavier nuclei, ab initio calculations of nuclear shape may be employed. A recent example illustrating the potential of this approach is the calculation of the ratio $r_{CC}/r_C$ in $^{26}$Mg with a relative uncertainty of $0.3\%$~\cite{he2026taming}.
Therefore, it is promising and desirable to extend such calculations to more nuclei in this region.
\subsection{\label{Sec: Screen} Electron Screening}

So far, we have considered muonic atoms as simple bound states devoid of electrons, which is only true for dilute light gases~\cite{1999-Kirch, 2016-PiMass,2023-NePRL, 1989-refill}. 
Relaxing this assumption leads to modifications of the muonic energy levels due to electron screening effects. These shifts are sub ppm for $1s$ states, so they do not interfere with radius extractions (see e.g.~\cite{2026-hybrid}); however, they reach several ppm for the $3d-2p$ centroid~\cite{beltrami1986new}, necessitating a thorough understanding to make this proposed experiment feasible across a range of elements.

For a given occupation number and electron configuration, the shift of a transition compared to a muonic atom without electrons can be reliably calculated using mean field methods~\cite{1984-Screening} or more sophisticated methods~\cite{2023-NePRL}. However, experimental data are necessary to explore the allowable parameter space of electron configurations.
Given that the magnitude of the shift increases steeply with the principal quantum number (scaling as $n^6$ for low-$n$ circular transitions), the electronic configuration can be inferred from transition energies involving higher excited states measured with a moderate accuracy goal compared to that of the $3s-2p$ manifold~\cite{1984-Screening}.
In this context, the recent introduction of microcalorimeters for energy measurements in exotic atoms is highly beneficial, as their broadband nature enables the simultaneous measurement of several weak, high-lying transitions~\cite{2023-NePRL, Eizenberg2026Be}. Thus, we propose to determine the electron configuration while simultaneously searching for new physics by measuring additional transitions, as successfully demonstrated in muonic neon~\cite{2023-NePRL} and silicon~\cite{1984-Screening}.

\subsection{\label{Lineshape}Line-shape considerations}

\begin{figure}[tbp]
    \centering
    \includegraphics[width=0.95\linewidth, trim=0 3mm 0 2mm, clip]{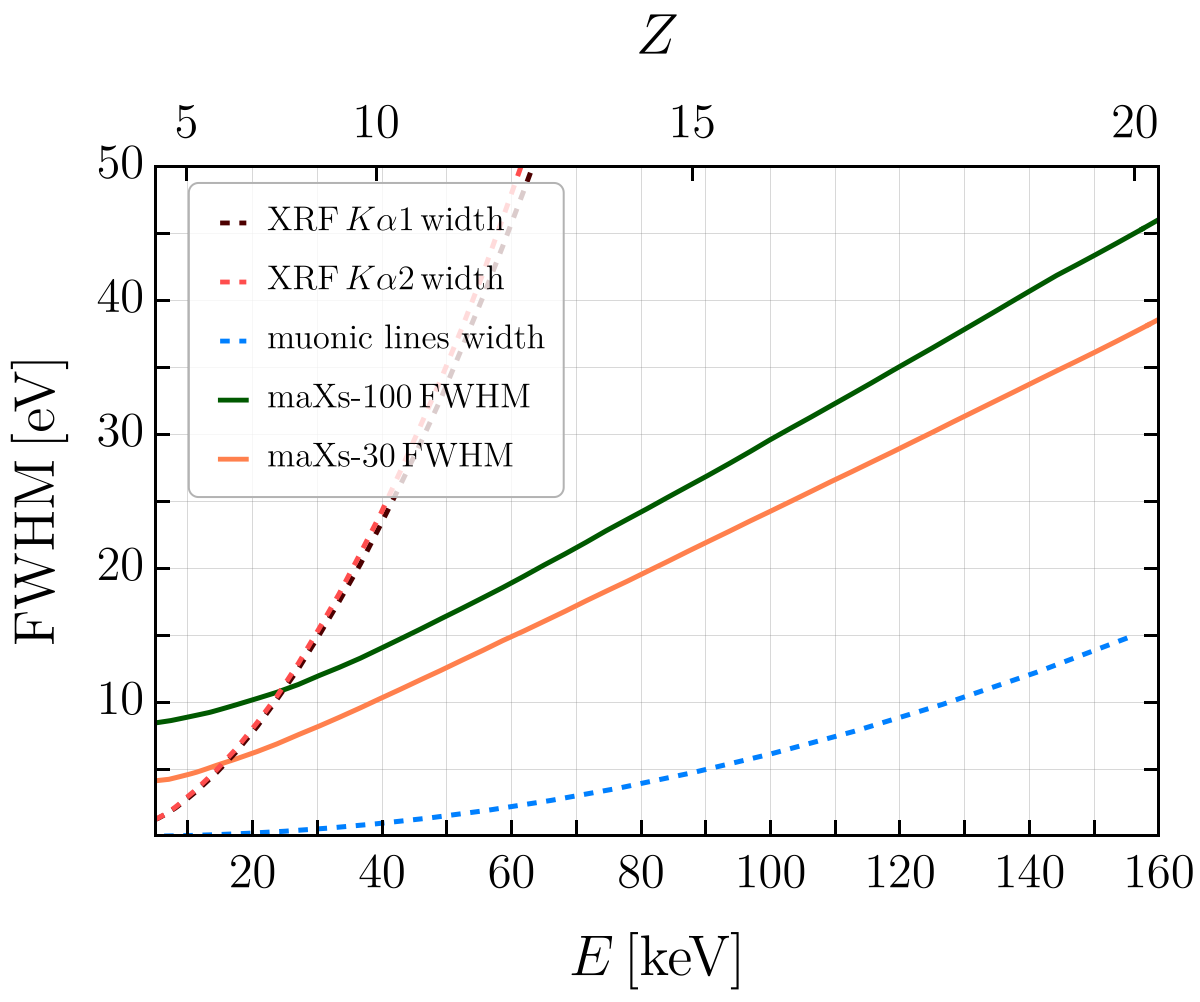}
    \caption{Estimated attainable resolution as function of energy for two considered detectors, along with natural linewidth of muonic and XRF lines.
    The top axis indicates the atomic numbers whose $3d-2p$ manifold falls within the bottom axis energy.}
    \label{fig:MMC FWHM and natural widths}
\end{figure}

Microcalorimeters are advantageous not only because they enable the concurrent measurement of multiple transitions, helping to control for electron-screening effects, but also because they provide high energy resolution. The highest resolutions, on the order of a few eV full width at half maximum (FWHM), are achieved with detectors employing thin absorbers~\cite{2012-MMC, 2021-TES, 2021-6eV}. However, this design choice entails a reduction in stopping power at higher photon energies. Consequently, detector selection involves optimizing spectral resolving power while maintaining adequate overall detection efficiency.
We illustrate this by considering two state-of-the-art detectors employed in muonic atom research: the maXs-30, which has a $20~\mu$m absorber, a baseline resolution of $5\,$eV, and a quantum efficiency above $10\%$ up to $70\,$keV, and the maXs-100, equipped with a $100~\mu$m absorber, a baseline resolution of $8\,$eV, and a quantum efficiency above $10\%$ up to $160\,$keV~\cite{2024-Unger, unger2025high, Weber2022MMC}. 

The energy dependence of the instrumental energy resolution for the two detectors is taken from Refs.~\cite{2024-Unger, unger2025high} and is shown in Fig.~\ref{fig:MMC FWHM and natural widths}, along with the negligible natural linewidths of the relevant muonic transitions. At room temperature, Doppler broadening in gaseous targets and phonon-induced broadening in solid targets are both significantly smaller than the natural linewidth and are therefore neglected. The dominant additional broadening mechanism is attributed to Coulomb explosion in molecular gasses, estimated to be on the order of $10^{-4}$ of the transition energy~\cite{2000-Coulomb}. This contribution exceeds the natural linewidth but remains smaller than the detector resolution. Consequently, the overall experimental resolution is expected to be primarily determined by the intrinsic resolution of the employed detector.

Although the $3d-2p$ transitions are the most intense, photons from all allowed transitions between the $n=3$ and $n=2$ manifolds are emitted during the cascade process. Dedicated measurements in light exotic atoms suggest that it is reasonable to assume that the fine structure (depending on quantum number $j$) and the magnetic dipole hyperfine structure (depending on quantum number $F$) are statistically populated in the cascade~\cite{1985-LiHFS, 1984-12C, beltrami1986new, 1982-3d-2p}, allowing the calculation of the relative magnitudes of the peaks as inputs for the fit.
However, the angular-dependence (depending on quantum number $l$) of the muonic cascade is highly non-trivial. In light of this, we argue that any reasonable lineshape parameterization should include at least one additional free parameter corresponding to the ratio between the magnitudes of the $3d-2p$ and $3p-2s$ transitions. To facilitate this, the Lamb Shifts (differences between the 2s and 2p states) must be well resolved by the detector, suggesting the choice of $5<Z$~\cite{1985-LS, 2018-LAserLS}.

The balance between having a sufficiently high $Z$ for the lineshape to consist of separate components and a low enough $Z$ for QED and nuclear corrections to be well understood indicates that nuclei with atomic numbers $Z = 6\text{–}12$ are the most suitable candidates for the highest-precision measurements, provided that an appropriate calibration strategy can be employed.

\subsection{\label{Calibration}Calibration}

The capability to convert measured signal amplitudes into corresponding energy values is achieved by recording the response of the system to an unknown transition simultaneously with the responses to several well-characterized reference lines of precisely known energies.
The energy difference between the calibration line and the line under investigation must also be minimized, as large separations increase systematic uncertainties arising from nonlinear processes, such as those intrinsic to  detectors~\cite{2006-TES, 2018-MMC, 2020-ThPRL} and analog-to-digital conversion~\cite{rodrigues2026measurements}.

\begin{figure}[tbp]
    \centering
    \includegraphics[width=0.95\linewidth, trim=0 2mm 0 2mm, clip]{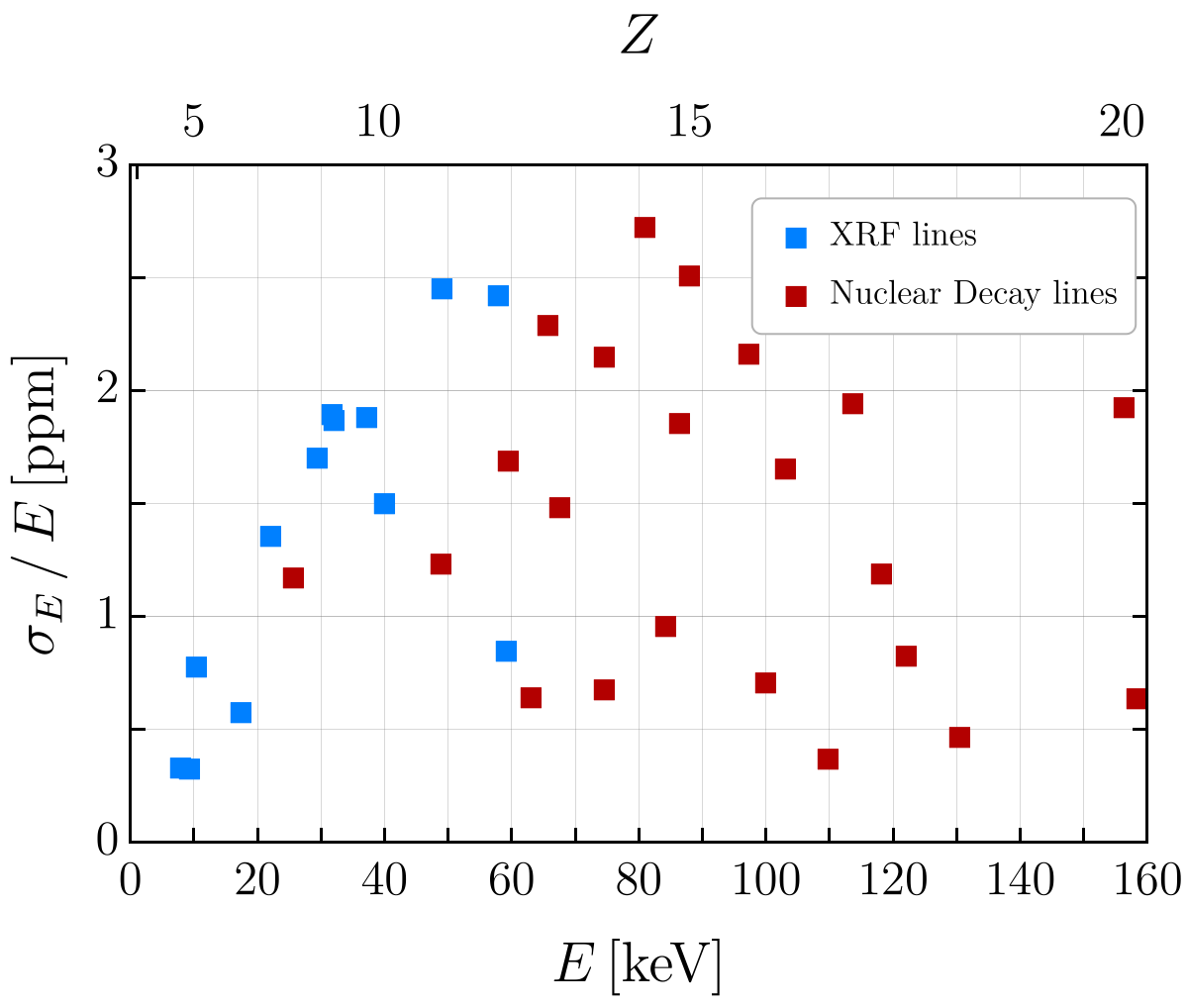}
    \caption{Fractional uncertainty in existing reference lines relevant to this work. The top axis and dashed vertical lines indicate the atomic numbers whose $3d-2p$ manifold falls within the bottom axis energy.
    }
    \label{fig:calibration lines}
\end{figure}

We consider two classes of calibration sources. The first consists of $K\alpha$ characteristic x-ray fluorescence (XRF) lines, whose energies are precisely known from measurements using crystal spectrometers. Figure~\ref{fig:calibration lines} shows the fractional uncertainties reported for these lines up to $60$\,keV~\cite{1997-Holzer, 2003-NISTxray, 2017-Mendenhall, 2019-Mendenhall}, highlighting their significance as references for light muonic atoms. However, as shown in Fig.~\ref{fig:MMC FWHM and natural widths}, above $40$\,keV, the intrinsic linewidth of these transitions exceeds twice the detector resolution~\cite{10.1063/1.555595}, limiting their usefulness as high-precision calibration standards at higher energies.

The second category of calibration sources consists of radioactive isotopes that emit $\gamma$ rays with well-measured energies~\cite{2000-Helmer, rodrigues2026measurements}. These sources have negligible intrinsic and Doppler-broadened linewidths but present their own challenges. Common calibration sources produce parasitic emissions that add unwanted spectral features, which may overload the detector. In addition, several well-measured emitters, such as $^{161}$Tb, have short half-lives and must be produced shortly before the measurement campaign.

Figure~\ref{fig:calibration lines} shows that the transition of interest in each candidate element is close to one or more well known reference lines, allowing the calibration uncertainty to be maintained under $1.5\,$ppm in most cases.
Although some systems are expected to pose greater challenges than others, we anticipate that if the uncertainty value in existing calibration sources becomes the primary limiting factor, the metrology community can improve these, as demonstrated for XRF sources~\cite{2017-Mendenhall, 2019-Mendenhall, 2022-NISTDCS} and gamma-ray lines~\cite{2003-gams,2020-60keV, rodrigues2026measurements}.

\subsection{\label{Statistics}Beamtime estimates}

\begin{figure}[tbp]
    \centering
    \includegraphics[width=0.95\linewidth, trim=0 4mm 0 2mm, clip]{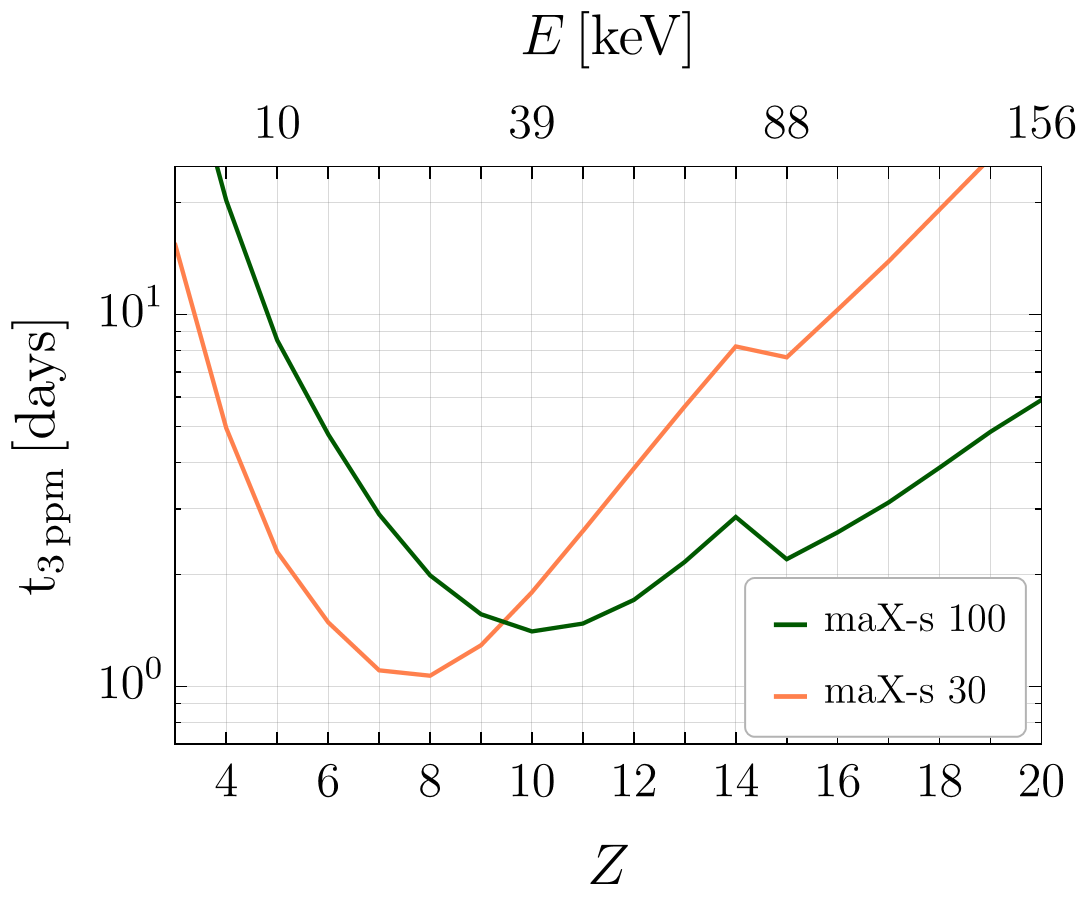}
    \caption{
    Estimated time to reach a competitive accuracy for different elements with two exemplary detectors.
    The top axis indicates the centroid energy of the $3d-2p$ manifold for a certain $Z$.
    }
    \label{fig:time for 3 ppm}
\end{figure}

As illustrated in Fig.~\ref{fig:bounds on new physics}, an accuracy level of 3\,ppm or better is required to access previously unexplored regions of the BSM parameter space. Achieving this with a detector whose resolving power is on the order of several thousand (see Fig.~\ref{fig:MMC FWHM and natural widths}) requires the collection of approximately $10^4$ events. Assuming a continuous muon rate of $10^4$\,s$^{-1}$, a geometric collection efficiency of $10^{-4}$~\cite{Eizenberg2026Be}, and a quantum efficiency above 10\%, the required acquisition time is on the order of days.
A refined time estimate, based on the quantum efficiency curves from Refs.~\cite{2024-Unger, unger2025high, Weber2022MMC} and using similar data-quality cuts as in Ref.~\cite{Eizenberg2026Be}, is illustrated in Fig.~\ref{fig:time for 3 ppm}.
The results indicate that a low-energy detector is optimal for measuring light muonic atoms up to $Z = 9$, while detectors with thicker absorbers are preferable for heavier systems.
Elements with $Z = 6$–$12$ are the best candidates for achieving a $1\,\mathrm{ppm}$ measurement, which would take $10-15$ days.

\section{Conclusions}
Based on the discussion above, we identify two promising candidates for this study. 
The first is $^{16}$O, whose relevant transitions occur around $24.9\,$keV, between the Ag K$\alpha_1$ line at $22.162917(30)\,$keV, known to $1.4$ ppm~\cite{Deslattes1985, 2003-NISTxray}, and the $25.651358(30)\,$keV line emitted by $^{161}$Tb, which is known to be $1.2\,$ppm~\cite{2000-Helmer}.
This nucleus has a low charge number, simplifying QED calculations, and a closed-shell configuration that makes nuclear structure calculations more manageable.
Moreover, dedicated studies suggest that at least up to a pressure of $1.4$~bar, electrons are absent for low lying transitions in muonic oxygen~\cite{2016-PiMass}.
An experimental configuration with a low-energy microcalorimeter can measure to $3\,$ppm within one day (Fig.~\ref{fig:time for 3 ppm}) and so can reach $1\,$ppm within $10\,$days.
A challenging aspect of this measurement is stopping enough muons at sufficiently low pressure and using a calibration source with a half life of 7 days.
Notwithstanding the challenges, such a measurement can probe the region between $m_X=0.1$\,MeV and $m_X=1$\,MeV that is not constrained by other experiments (see Fig.~\ref{fig:bounds on new physics}). This measurement could also decouple new physics effects from fundamental constant determinations by covering the area currently excluded by hydrogen lamb shifts.

Although light target nuclei are preferable from both theoretical and experimental standpoints, they do not provide competitive sensitivity to potential new physics above $m_X\sim 3\,$MeV (see Fig.~\ref{fig:bounds on new physics}). To explore this energy range, heavier target nuclei are necessary, even at the cost of increased uncertainties. 
To illustrate this point, we consider a low-pressure $^{40}$Ar target, where electron dynamics are well understood~\cite{2025-Ar}. The relevant nuclear transitions lie near $128\,$keV, between the $122.06065(12)\,$keV line of $^{57}$Co, whose energy is known to $1.0$\,ppm~\cite{2000-Helmer}, and the $130.52293(6)\,$keV line of $^{169}$Yb, known to $0.5$\,ppm~\cite{2000-Helmer}. 
The primary experimental difficulty is accumulating sufficient statistics due to reduced resolution and quantum efficiency (see Fig.~\ref{fig:time for 3 ppm}). Consequently, we envisage a $\sim12$ day measurement targeting a relative precision of $2\,$ppm. 
On the theoretical side, a better understanding of nuclear polarization (Fig.~\ref{fig:QED and NP uncertainty}) and shape (Fig.~\ref{fig:FNS uncertainty}) are required. Despite these challenges, such a measurement would enable further exploration of the BSM parameter space above $3$\,MeV, as shown in Fig.~\ref{fig:bounds on new physics}.

In summary, precision spectroscopy of the $3d\!\to\!2p_{3/2}$ manifold in muonic atoms offers a clean and complementary route for testing short-range, spin-independent interactions between muons and protons mediated by new bosons with masses in the MeV range.
We show that achieving up to $\sim 1$ ppm accuracy is realistic with microcalorimeter detectors and reasonable muon rates, provided that electron screening is experimentally constrained using additional higher-lying lines and that calibration and line-shape systematics are controlled. On the theory side, we highlight a need to calculate specific missing higher-order QED contributions, develop nuclear polarization theory for contact-free transitions beyond leading order, and complement data-driven estimations of dipole polarizability with \textit{ab initio} calculations. 
We show that the dominant theory uncertainty arises from our limited knowledge of nuclear shape, which can be improved using \textit{ab initio} calculations. These considerations motivate near-term measurements in light systems such as $^{16}$O for the most stringent precision goals and a complementary heavier-nucleus program (e.g., $^{40}$Ar) to extend sensitivity toward higher mediator masses.

\begin{acknowledgments}
We thank Shikha Rathi, Vladimir Yerokhin, Krzysztof Pachucki, Yotam Soreq, Ulrich Jentchura, and Tim Egert for their valuable suggestions.
\end{acknowledgments}

\bibliography{references.bib}

\end{document}